\documentclass[aps,pra,reprint,groupedaddress]{revtex4-1}

\usepackage{amsmath}
\usepackage{amssymb}
\usepackage{graphicx}
\usepackage[usenames,dvipsnames,svgnames]{xcolor}
\usepackage{bm}
\usepackage{epstopdf}
\usepackage[colorlinks,citecolor=blue]{hyperref} 

\newcommand{\rd}{d}
\newcommand{\ri}{i}

\newcommand{\p}{\partial}
\newcommand{\avg}{\overline}
\newcommand{\gru}{\rho}

\begin{document}

\title{Quantum dynamics in potentials with fast spatial oscillations}

\author{Viktor Novi\v{c}enko}
\email[]{viktor.novicenko@tfai.vu.lt}
\homepage[]{http://www.itpa.lt/~novicenko/}
\affiliation{Institute of Theoretical Physics and Astronomy, Vilnius University,
Saul\.{e}tekio Ave.~3, LT-10257 Vilnius, Lithuania}

\author{Julius Ruseckas}
\email[]{julius.ruseckas@tfai.vu.lt}
\homepage[]{http://web.vu.lt/tfai/j.ruseckas/}
\affiliation{Institute of Theoretical Physics and Astronomy, Vilnius University,
Saul\.{e}tekio Ave.~3, LT-10257 Vilnius, Lithuania}

\author{Egidijus Anisimovas}
\email[]{egidijus.anisimovas@ff.vu.lt}
\affiliation{Institute of Theoretical Physics and Astronomy, Vilnius University,
Saul\.{e}tekio Ave.~3, LT-10257 Vilnius, Lithuania}

\date{\today}

\begin{abstract}
We consider quantum dynamics of systems with fast spatial modulation of the Hamiltonian. 
Employing the formalism of supersymmetric quantum mechanics and decoupling fast and slow 
spatial oscillations we demonstrate that the effective dynamics is governed by a 
Schr\"{o}dinger-like equation of motion and obtain the expression of the resulting 
effective Hamiltonian. In particular, we show that there exists an attractive effective 
potential even in the case when the oscillating potential averages to zero.
\end{abstract}


\maketitle

\section{Introduction}

The idea to simplify the analysis of a physical problem by taking into account the 
presence of substantially different characteristic scales (spatial, temporal or other) 
is a ubiquitous and powerful one.  Far-reaching examples include the Born-Oppenheimer 
approach \cite{BornOppen,Kolsos}, 
based on decoupling of fast and slow vibration modes, and 
Floquet engineering \cite{Eckardt17RMP}, 
which fruitfully exploits the idea to realize effective Hamiltonians with desired 
properties by applying a time-periodic driving to a controllable 
physical system \cite{GrossBloch17}. 
In the limit when the periodic driving sets the dominant
frequency scale (in comparison to the internal dynamics of the system) the resulting
(stroboscopic) dynamics is described by the time-averaged driven Hamiltonian. Already
this rather simple result has led to numerous insights and ground-breaking experimental
schemes \cite{Aidelsburger:2013,MiyakeEtAl13,Atala14,Aidelsburger14NP,Kennedy15,TaiEtAl17}.
When the driving sets the largest but not overwhelming frequency scale, the 
resulting dynamics may be captured by a systematic inverse-frequency 
expansion \cite{Goldman2014,Eckardt2015,Novicenko2017},
which also offers opportunities for quantum engineering of physically interesting 
model Hamiltonians \cite{Jotzu2014,GrushinEtAl14}

In the present contribution, we look at the complementary facet of \emph{spatial},
rather than temporal, modulation of the Hamiltonian. We show that in the limit of 
rapid oscillations of the potential, the effective dynamics is governed by a
Schr\"{o}dinger-like equation of motion. The main contribution to the effective 
potential featured in this equation is proportional to the square of the envelope 
function that modulates the rapid oscillations of the true potential. Moreover,
this contribution comes with a negative sign, i.e.\ there exists an attractive
effective potential even when the oscillating potential averages to zero.
The effective
Schr\"{o}dinger equation is solved by a smoothed wave function which accurately 
approximates the overall shape of the true wave function but excludes its rapid
small-scale oscillations. The obtained description is relevant to describe quantum
dynamics in potentials formed by interfering laser beams. Here, the resulting  
intensity distributions, which define the potentials felt by ultracold atoms,
typically combine rapid variations on the scale of the wavelength with slow 
modulation due to the shape of the beams \cite{Windpassinger2013RPP}.

Interestingly, in the derivation of the effective Schr\"{o}dinger equation we
benefited from an approach based on supersymmetric quantum mechanics (SUSY QM).
SUSY QM is a generalization of the factorization method commonly used for the harmonic 
oscillator. It was first introduced as a model to study non-perturbative symmetry breaking 
in supersymmetric field theories \cite{Witten1981}. Later it was realized that SUSY QM
is an interesting field in its own right, and the ideas of supersymmetry have been profitably 
applied to many quantum mechanical problems; see 
Refs.~\cite{Cooper1995,SUSYBook2001,SUSYBook2018} for books and reviews.

Our paper is organized as follows: In Sec.~\ref{sec:eff-ham} we derive the expression 
for the effective potential in a system with fast spatial potential modulation. To verify 
the validity of the effective Hamiltonian, in Sec.~\ref{sec:numerical} we compare numerically
calculated eigenfunctions of the original Hamiltonian with the eigenfunctions obtained using 
the effective Hamiltonian. 
In Sec.~\ref{sec:application} we discuss applications to potentials for ultracold atoms 
formed by interfering laser beams.
Finally, in Sec.~\ref{sec:concl} we summarize our findings.

\section{Dynamics in a spatially modulated potential}
\label{sec:eff-ham}

We study the motion of a one-dimensional quantum particle of mass $m$ in a static spatially 
modulated potential $V(x)$ whose shape can be represented as a combination of a slowly varying 
envelope function $\Phi (x)$ with vanishing limiting values
\begin{equation}
\label{eq:philimit}
  \lim_{x\rightarrow \pm \infty} \Phi(x) = 0,
\end{equation}
and a periodic function $v (kx) = v (kx + 2\pi)$. The period of the rapid oscillations 
of the potential is given by $2\pi/k$ and by assumption must be much smaller than 
the characteristic length scale of the envelope function $\Phi (x)$. The dynamics of 
the particle is described by the Schr\"odinger equation
\begin{equation}
  \ri\hbar\frac{\partial}{\partial t}\psi = \hat{H}\psi,
\end{equation}
with the Hamiltonian
\begin{equation}
  \hat{H} = -\frac{\hbar^2}{2m}\frac{\partial^2}{\partial x^2} + V(x).
\end{equation}
Our goal is to derive an effective Hamiltonian that approximates the ensuing dynamics for 
large but finite $k$, and becomes exact in the limit $k\rightarrow \infty$. 
The potential-energy term is taken of the form
\begin{equation}
\label{eq:vvv}
  V (x) =  k v \left( kx \right) \Phi (x) + \gru (x).
\end{equation}
Here, $v (s)$ is a periodic function that averages to zero over a period
\begin{equation}
\label{eq:zeroavg}
  \left\langle v \right\rangle= \frac{1}{2\pi} \int_0^{2\pi} v(s) \, \mathrm{d}s = 0,
\end{equation}
and in Eq.~(\ref{eq:vvv})
is additionally scaled by a factor $k$, that is, the increasing frequency of spatial 
oscillations is complemented by increasing amplitude. As further explained below, 
if this was not done, in the limit $k \to \infty$ the oscillatory potential 
$v(kx) \Phi (x)$ would average out.
The presence of a finite potential background is taken into account by a 
separate term $\gru (x)$, which is also required to vanish for $|x| \to \infty$, 
cf.~Eq.~(\ref{eq:philimit}).
To be able to work with dimensionless quantities, we identify 
a characteristic length scale $\ell$ and measure coordinates in units $\ell$,
wavenumbers in $\ell^{-1}$, energies in units $\hbar^2 / (2m\ell^2)$ and time in 
$2m\ell^2/\hbar$. 
The dimensionless Schr\"{o}dinger equation reads
\begin{equation}
  \ri \frac{\p}{\p t} \psi
  = \left\{ -\frac{\p^2}{\p x^2} 
  + k v\left( kx\right) \Phi(x) + \gru (x) \right\} \psi (x,t).
\label{main}
\end{equation}
%
We note that a related problem, as a specific case, was studied in~Ref.~\cite{Ruiz2017}, 
where ponderomotive dynamics was derived as an expansion with respect to the inverse 
wavenumber $k^{-1}$. However, here we are interested in the regime of large oscillation 
amplitude [in comparison to the particle recoil energy $\hbar^2/(2m \ell^2)$]. 
This regime can not be directly covered by the formalism developed in Ref.~\cite{Ruiz2017}.

\subsection{Derivation of the effective Hamiltonian using SUSY QM formalism}

We derive the effective Hamiltonian using an approach based on the formalism of 
SUSY QM (see for example~\cite{Cooper1995}) and 
the subsequent application of the `averaging' theorem~\cite{burd07,sand07}, which 
is a versatile tool that allows to eliminate rapidly oscillating terms in broad 
classes of first-order differential equation sets. (An alternative derivation is 
included as an Appendix.) To proceed, we write the Hamiltonian in a factorized form,
\begin{equation}
  \hat{H} = \hat{A}^{\dag}\hat{A},
\end{equation}
where
\begin{equation}
  \hat{A} = \frac{\p}{\p x} + W(x),
\end{equation}
and we introduced the superpotential $W(x)$ as a solution of the differential equation
\begin{equation}
  \frac{\rd}{\rd x} W(x) = W^2(x) - k v \left( kx \right) \Phi(x) - \gru (x),
\label{sup}
\end{equation}
known as the Riccati equation.
Then we can replace the Schr\"odinger equation~(\ref{main}) by a pair of 
first-order differential equations 
\begin{subequations}
\label{set} 
\begin{align}
  \frac{\p}{\p x} \psi(x,t) & = \varphi(x,t) - W(x)\psi(x,t), \label{set_psi}\\
  \frac{\p}{\p x} \varphi(x,t) & = -\ri \frac{\p \psi(x,t)}{\p t} 
  + W(x)\varphi(x,t), 
\label{set_phi}
\end{align}
\end{subequations}
which govern the wave function $\psi(x,t)$ and an auxiliary function 
$\varphi(x,t) = \hat{A}\psi(x,t)$.

To be fully compliant with the requirements of the averaging theorem one should replace the 
partial derivatives $\p/\p x$ with the full derivatives $\rd / \rd x$. Formally, this can be 
done by discretizing the time interval into small steps of duration $\Delta t$, and introducing
the notation $\psi_{(m)}(x) \equiv \psi(x,m\Delta t)$ and 
$\varphi_{(m)}(x) \equiv \varphi(x,m\Delta t)$. Then the term $\p \psi(x,t)/\p t$ can be 
approximated by a finite difference 
$\p \psi(x,m\Delta t)/\p t \approx \left[ \psi_{(m+1)} - \psi_{(m)}\right]/\Delta t$.
As a consequence, Eqs.~(\ref{set}) turn into set of differential equations for the 
functions $\psi_{(m)}(x)$ and $\varphi_{(m)}(x)$ formulated in terms of full derivatives 
with respect to the single variable $x$. However, to keep the notation simple we retain
the partial derivatives having in mind that their presence poses no practical problems.

The three functions $W$, $\psi$ and $\varphi$ comprise the new dynamical variables 
and are governed by the set of first-order nonlinear differential equations (\ref{sup}) 
and (\ref{set}). Before proceeding to the averaging procedure, let us review the scaling
of the potential energy, which we choose to write as $k v(kx)\Phi(x)$. At this stage we are 
in a position to see that -- in the absence of the additional scaling by the factor $k$ -- 
the term $v(kx) \Phi (x)$ would be eliminated from Eq.~(\ref{sup}) by the averaging procedure 
as a rapidly oscillating term. In the presence of the scaling factor $k$, the potential
energy $k v(kx)\Phi(x)$ becomes formally divergent in the $k \to \infty$ limit and must be 
cast into a manageable form. To achieve this aim, we replace $W$ with a new variable
\begin{equation}
W^{\prime} = W + g(kx)\Phi(x),
\label{W_prime}
\end{equation}
where the function $g(s)$ is the zero-average antiderivative of the function 
$v(s)$, i.e.,
\begin{equation}
  g(s) = \int\limits_0^s v(s^{\prime}) \, \rd s^{\prime}
   - \frac{1}{2\pi} \int\limits_0^{2\pi} \int\limits_0^{s^{\prime\prime}} 
   v(s^{\prime})\, \rd s^{\prime} \, \rd s^{\prime\prime}.
\label{anti}
\end{equation}
The purpose of the transformation from $W$ to $W'$ can be elucidated by differentiating both 
sides of Eq.~(\ref{W_prime}) with the result
\begin{equation}
  \frac{d}{dx} W + k v (kx) \Phi (x) = \frac{d}{dx} W' - g (kx) \frac{d}{dx} \, \Phi (x),
\end{equation}
demonstrating the absorbtion of the term proportional to $k$. Thus, the new variables 
$W'$, $\psi$, and $\varphi$ obey the differential equations
\begin{subequations}
\label{set1} 
\begin{align}
  \frac{\rd W^{\prime}}{\rd x} & = \left[ W^{\prime} 
    - g \left( kx \right) \Phi(x) \right]^2 
    - \gru (x) + g(kx) \frac{\rd \Phi}{\rd x}, 
  \label{set1_w}\\
  \frac{\p \psi}{\p x} & = \varphi 
  -\left[ W^{\prime} - g \left(kx \right) \Phi(x) \right] \psi, \label{set1_psi}\\
  \frac{\p \varphi}{\p x} & = - \ri \frac{\p \psi}{\p t}
  + \left[ W^{\prime} - g \left(kx \right) \Phi(x) \right] \varphi . \label{set1_phi}
\end{align}
\end{subequations}
Here we consider continuous envelope functions $\Phi(x)$ so that the spatial
derivative of the envelope represented by the term $\rd \Phi/\rd x$ in
equation~(\ref{set1_w}) remains finite. The more general case of piecewise
continuous envelope functions is treated in the following Subsection~\ref{sec:piece-wise}.

Now one can apply the averaging theorem~\cite{burd07,sand07}. In general, the averaging 
theorem can be formulated as follows:  Let us consider the vector $\mathbf{Z}(x)$ 
obeying the differential equation of the form
\begin{equation}
  \frac{\rd \mathbf{Z}}{\rd x} = 
  \mathbf{F} \left( \mathbf{Z},kx,x,k \right),
\label{avr_form}
\end{equation}
where the vector field $\mathbf{F}$ is $2\pi$-periodic with respect to the second argument,
slowly depends (in comparison to the characteristic period $2\pi/k$) on third argument and
additionally depends on parameter $k$ represented by the fourth argument. One can 
introduce the averaged vector field as
\begin{equation}
  \avg{\mathbf{F}} \left( \mathbf{Z} ,x \right)
  = \frac{1}{2\pi} \int\limits_0^{2\pi} 
  \left[ \lim_{k\rightarrow \infty} 
   \mathbf{F}\left( \mathbf{Z},\vartheta,x,k \right) \right] \rd \vartheta .
\label{avr_field}
\end{equation}
According to the averaging theorem, the solution $\avg{\mathbf{Z}}$ of the differential equation
\begin{equation}
  \frac{\rd \avg{\mathbf{Z}}}{\rd x}
  = \avg{\mathbf{F}}\left( \avg{\mathbf{Z}},x \right)
\label{avr_form1}
\end{equation}
with the identical initial conditions $\mathbf{Z}(x_0) = \avg{\mathbf{Z}}(x_0)$
approximates the original solution 
$\mathbf{Z}(x) = \avg{\mathbf{Z}}(x)+\mathcal{O}\left(k^{-1}\right)$. 
Thus from Eqs.~(\ref{set1}) one finds 
\begin{subequations}
\label{set2} 
\begin{align}
  \frac{\rd \avg{W^{\prime}}}{\rd x} & = \avg{W^{\prime}}^2 
  + \langle g^2 \rangle \Phi^2(x) - \gru (x), 
  \label{set2_w}\\
  \frac{\p \avg{\psi}}{\p x} & = \avg{\varphi} -\avg{W^{\prime}}\, \avg{\psi},     
\label{set2_psi}\\
  \frac{\p \avg{\varphi}}{\p x} & = -\ri \frac{\p \avg{\psi}}{\p t}
  + \avg{W^{\prime}} \avg{\varphi}, \label{set2_phi}
\end{align}
\end{subequations}
where the averaged functions approximate the original functions as
\begin{equation}
  \left\lbrace \avg{W^{\prime}}, \avg{\psi}, \avg{\varphi} \right\rbrace 
  = \left\lbrace W^{\prime}, \psi, \varphi \right\rbrace
  + \mathcal{O}\left(k^{-1}\right).
\label{approx}
\end{equation}
We note here, that the use of two distinct notations for `averaging' is justified
by their distinct meanings: the brackets $\langle v \rangle$ refer to the mean 
value of a periodic function $v$ taken over its period, whereas the overline 
$\avg{\psi}$ refers to a smoothed approximation of an oscillatory function, 
cf.\ Fig.~\ref{fig_eigenf}.

Finally, transforming Eqs.~(\ref{set2}) back to a single second-order equation for 
the smoothed function $\avg{\psi}$, we obtain
%
\begin{equation}
  \ri \frac{\p}{\p t}\avg{\psi} 
  =-\frac{\p^2}{\p x^2} \avg{\psi}
  + V_{\rm eff}(x) \avg{\psi},
\label{main_eff1}
\end{equation}
with
\begin{equation}
  V_{\rm eff}(x) =
  - \langle g^2 \rangle \Phi^2(x) + \gru (x) .
\label{main_eff2}
\end{equation}
%
This result shows that the dynamics is governed by an effective equation of motion which
retains the form of the usual Schr\"{o}dinger equation with the effective Hamiltonian
\begin{equation}
\hat{H}_{\mathrm{eff}}=-\frac{\p^2}{\p x^2} + V_{\rm eff}(x),
\end{equation}
where the effective potential (\ref{main_eff2})
includes an attractive contribution proportional to the \emph{square} of the envelope 
function. In the hindsight, this is intuitively clear: if there is no background, 
i.e.\ $\gru (x) = 0$, the sign of the envelope 
function has no effect and only even powers of $\Phi (x)$ can contribute. 
Finally, let us stress that even though the wave function $\avg{\psi}$ approximates 
the original wave-function $\psi$ with the same accuracy as (\ref{approx}), i.e.\  
$\avg{\psi} = \psi+\mathcal{O}\left(k^{-1}\right)$, an analogous statement
\emph{does not hold} for either the kinetic energy or the potential energy calculated 
using $\avg{\psi}$. Only the total energy is well approximated by the quantum-mechanical 
expectation value of the effective Hamiltonian with the smoothed wave function $\avg{\psi}$.

\subsection{Effective potential for piecewise continuous envelope}
\label{sec:piece-wise}

Discontinuous envelope functions do not satisfy the requirement that the
characteristic length scale of the envelope function should be much larger than the 
period of the rapid oscillations. Thus the effective potential~(\ref{main_eff2}) is
valid only for continuous envelope functions $\Phi(x)$. In this subsection we will 
generalize our approach to include the case when the envelope is a piecewise
continuous function. In this situation, the averaging is still applicable, however,
at the points of discontinuity the smoothed functions $\avg{\psi}$ must obey
appropriate boundary conditions.

For each coordinate $x$ which is not a point of discontinuity, all of the
steps~(\ref{sup})--(\ref{set2}) can be repeated in exactly the same way as before.
However, the points of discontinuity should be considered separately. At every
point of discontinuity $x_0$ the superpotential $W(x)$ must remain continuous
for any value of $k$ [cf.\ Eq.~(\ref{sup})]. Using Eq.~(\ref{W_prime}) for the
transformed superpotential $W'$ we obtain
\begin{equation}
  \lim_{\varepsilon\rightarrow +0}
  [W^{\prime}(x_0+\varepsilon) - W^{\prime}(x_0-\varepsilon)]
  = g(\varphi_{0})\Delta\Phi(x_0),
  \label{pw_super}
\end{equation}
where $\varphi_0 =k x_0$ is the phase of the oscillating function $g(kx)$
at the point of discontinuity $x_0$ and
$\Delta\Phi(x_0)=\lim_{\varepsilon\rightarrow +0}
[\Phi(x_0+\varepsilon) - \Phi(x_0-\varepsilon)]$.
This result indicates that at each point of discontinuity the transformed
superpotential exhibits a step given by the product of the step of the envelope 
function and the local value of the periodic antiderivative $g (kx)$. As a special
case, the transformed superpotential may remain continuous if $g(kx_0) = 0$.

We must now require that $k$ grows to infinity in discrete steps, i.e.\ by assuming 
a sequence of monotonically growing values for which the phase $\varphi_0$ remains 
the same (modulo $2\pi$).
Then the condition~(\ref{pw_super}) will also hold for averaged superpotential
$\avg{W^{\prime}}(x)$, and can be satisfied by including terms proportional to 
the Dirac delta 
function $\Delta\Phi(x_0)g(\varphi_0)\delta(x-x_0)$ to the equation~(\ref{set2_w}). 
If there are several points of discontinuity $\{x_n\}$ and it is possible to find 
values of $k$ that keep the phases $\varphi_n$ constant, the effective 
potential~(\ref{main_eff2}) becomes
\begin{align}
  V_{\rm eff}(x) = &-\langle g^2 \rangle \Phi^2(x) + \gru (x) \nonumber\\
                   &-\sum_{n}\Delta\Phi(x_n)g(\varphi_n)\delta(x-x_n).
  \label{v-eff-2}
\end{align}
We see that step discontinuities in the envelope function translate into 
delta-function singularities in the effective potential. As a special case, 
these singularities may be absent, if $g (\varphi_n) = 0$.

\section{Numerical examples}
\label{sec:numerical}

\begin{figure}
\centering\includegraphics[width=85mm]{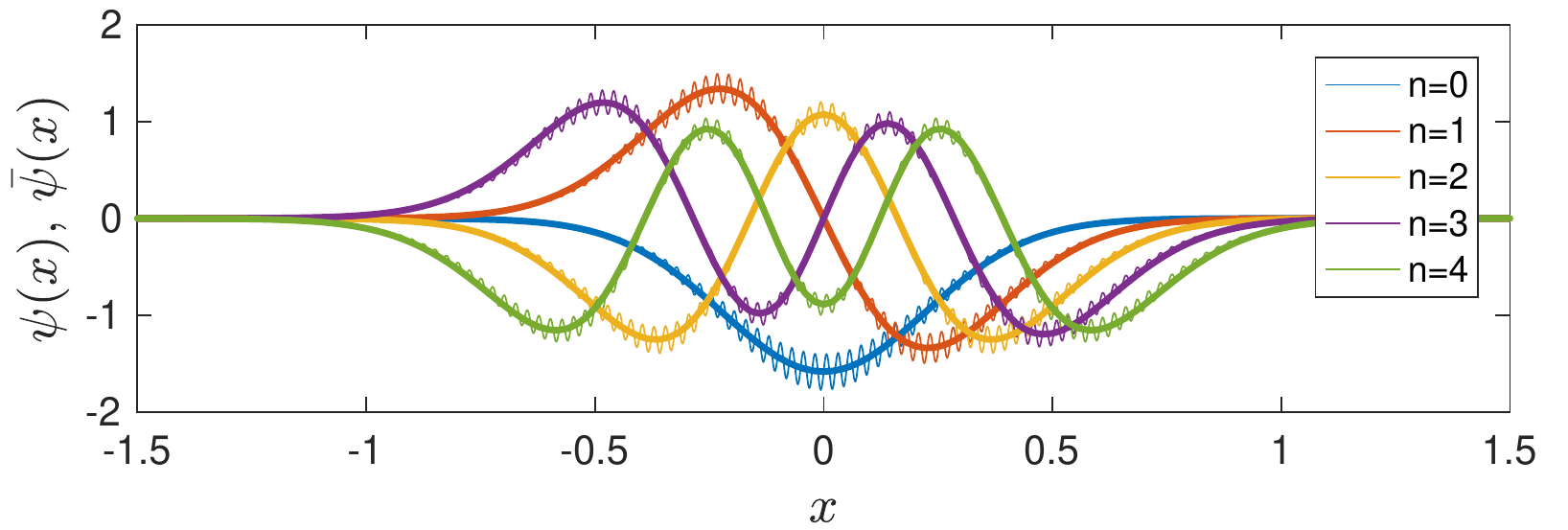}
\caption{\label{fig_eigenf} (Color online) The five lowest-energy states calculated
numerically using the original Hamiltonian~(\ref{main}) (thin oscillating curves)
and found analytically for the effective Hamiltonian~(\ref{pt}) (thick curves). 
The ground-state wave function is nodeless, and the excited states can be identified
by the number of nodes. The parameters are set to $k = 250$ and $a = 2\sqrt{210}$ 
such that $\lambda = 20$ is an integer number.}
\end{figure}

\begin{figure}
\centering\includegraphics[width=85mm]{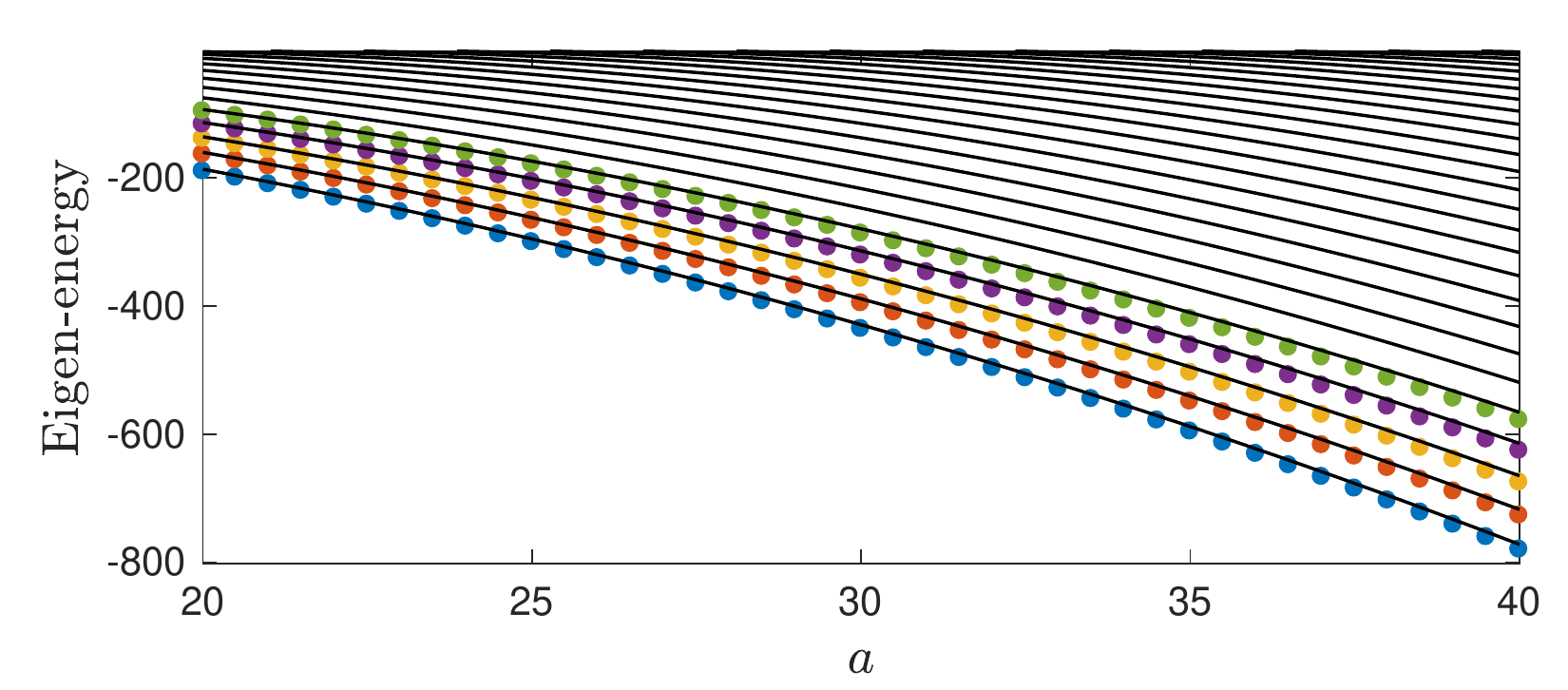}
\caption{\label{fig_ener} (Color online) The dependence of the lowest eigenenergies 
of the bound states on the depth of the potential well $a$ for fixed $k = 250$. Solid 
curves are drawn from Eq.~(\ref{energy}), while the circles are calculated numerically 
using the original Hamiltonian~(\ref{main}).}
\end{figure}

To verify the validity of the effective description we perform a numerical calculation 
of the eigenfunctions for the original Hamiltonian~(\ref{main}) and compare them with 
the eigenfunctions obtained using the effective Hamiltonian~(\ref{main_eff1}). We treat
both cases of continuous and piecewise continuous envelope functions and use the 
effective potentials given, respectively, by Eq.~(\ref{main_eff2}) and Eq.~(\ref{v-eff-2}).

Starting with the simpler case of a smooth envelope, the rapid potential oscillations 
are taken to have a harmonic shape, thus, $v(s) = \cos(s)$ and its antiderivative 
$g(s) = \sin(s)$ with $\langle g^2 \rangle = 1/2$. For the sake of convenience, the 
envelope function is described by $\Phi(x) = a \,\mathrm{sech}(x)$. Then the potential
well has a characteristic width of the order of unity, and the effective Schr\"{o}dinger
equation for the averaged wave function
\begin{equation}
  \bar{E} \bar{\psi} (x)
  = \left[ -\frac{\rd^2}{\rd x^2} - \frac{a^2}{2} \operatorname{sech}^2(x)\right] 
  \bar{\psi} (x),
\label{pt}
\end{equation}
is solvable analytically. The eigenvalue problem (\ref{pt}) is known as the 
P\"{o}schl-Teller problem \cite{PT1933}, and the spectrum of the bound states is given by
\begin{equation}
  \bar{E}_n = - \left(\lambda - n\right)^2.
\label{energy}
\end{equation}
Here, the parameter $\lambda = \left(\sqrt{1+2a^2}-1\right)/2$ is a function of the
depth $a$ and its integer part $\left\lfloor \lambda \right\rfloor$ (i.e., the largest 
integer smaller than or equal to $\lambda$) is equal to the number of bound states
plus one. The index $n = 0,1,\ldots,\left\lfloor \lambda\right\rfloor-1,\left\lfloor
\lambda\right\rfloor$ labels the bound states. 

The comparison of the first five eigenfunctions of the original and the effective 
Hamiltonians is represented in Fig.~\ref{fig_eigenf}. For convenience we choose 
the value of $a$ such
that $\lambda$ becomes an integer number. In that case the eigenfunctions 
$\bar{\psi}(x) \sim P_{\lambda}^{\lambda-n}(\tanh(x))$, where $P_{\lambda}^{\beta}(y)$ 
are the associated Legendre polynomials which are related to the ordinary Legendre 
polynomials $P_{\lambda}(y)$ as 
\begin{equation}
  P_{\lambda}^{\beta}(y) = (-1)^{\beta} 
  \left(1 - y^2\right)^{\beta/2} \frac{\rd^{\beta}}{\rd y^{\beta}} P_{\lambda}(y).
\label{legen}
\end{equation}
In Fig.~\ref{fig_ener}, we plot dependence
of the eigenenergies on the potential amplitude parameter $a$. As one can see in 
Figs.~\ref{fig_eigenf} and \ref{fig_ener}, the approximate expression obtained using 
the effective Hamiltonian shows good coincidence with results of numerical calculations.

\begin{figure}
\centering\includegraphics[width=0.9\columnwidth]{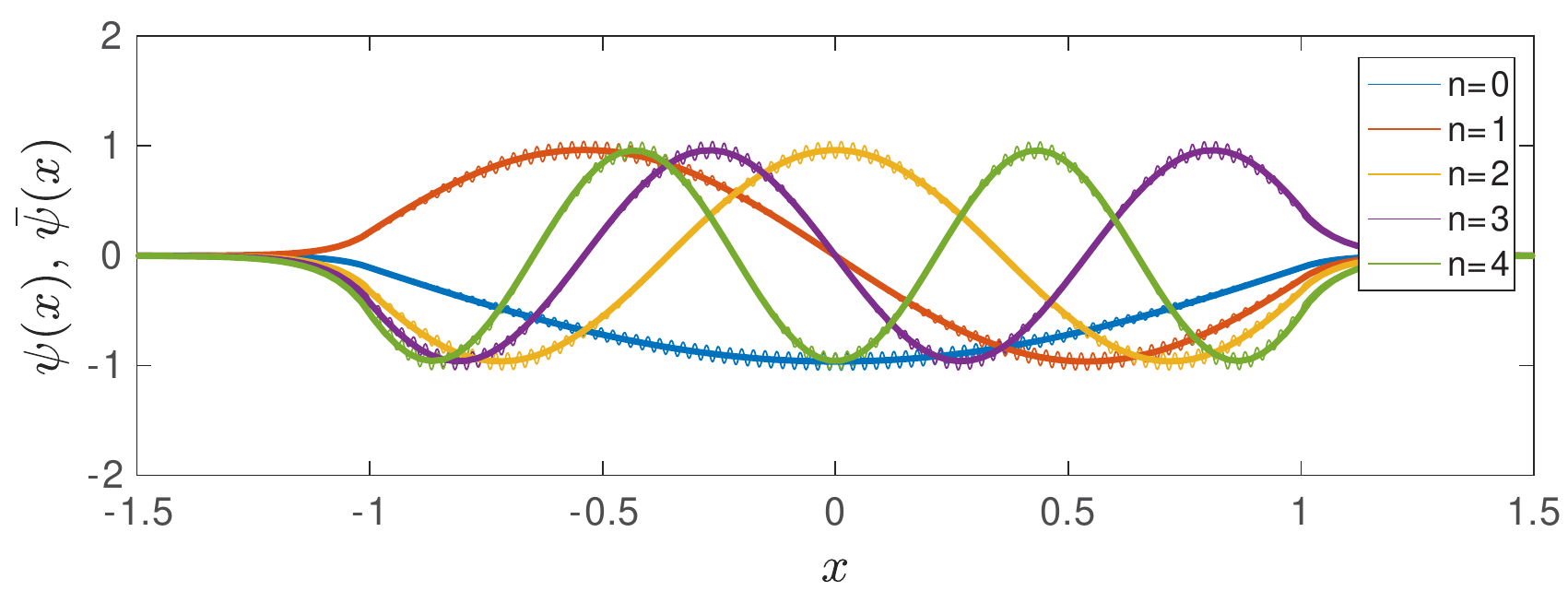}
\caption{\label{fig:pw_no_delta} (Color online) The case of piecewise
  continuous envelope~(\ref{pw_envel}). The five lowest-energy states
  calculated numerically using the original Hamiltonian~(\ref{main}) (thin
  oscillating curves) and found semi-analytically for the effective
  potential~(\ref{pw_pot1}) (thick curves). The parameters are set to $a = 20$
  and $k \approx 250$ is chosen in such a way that $g(\varphi_{-1,1})$
  vanishes.}
\end{figure}

To give an example of a piecewise continuous envelope, we consider the envelope 
function of the form of a square barrier:
\begin{equation}
\Phi(x)= \begin{cases}
a, & \textrm{ for }\left|x\right|<1,\\
0, & \textrm{ for }\left|x\right|>1.
\end{cases}
\label{pw_envel}
\end{equation}
According to Eq.~(\ref{v-eff-2}), the effective potential~(\ref{main_eff2})
reads
\begin{align}
  V_{\mathrm{eff}}(x) = &-\langle g^2\rangle\Phi^2(x)\nonumber\\
                        &+a[g(\varphi_{1})\delta(x-1) - g(\varphi_{-1})\delta(x+1)],
\label{pw_pot1}
\end{align}
with $\varphi_{\pm 1} = \pm 1 \cdot k$,
and features two delta-function terms at the points of discontinuity.

We perform numerical simulations with high frequency profile function
$v(s) = \cos(s)$ which gives $g(s) = \sin(s)$.
We choose an appropriate value of $k$ to ensure $\varphi_{\pm 1} = 0$, and the 
effective potential (\ref{pw_pot1})
has the shape of an ordinary symmetric quantum well of a finite depth $a^2/2$.
The comparison of the first five
eigenfunctions of the original and the effective Hamiltonian is presented in
Fig.~\ref{fig:pw_no_delta}. We note an excellent agreement of the two sets of
results, which demonstrates the applicability of the approach.

\begin{figure}
\centering\includegraphics[width=0.9\columnwidth]{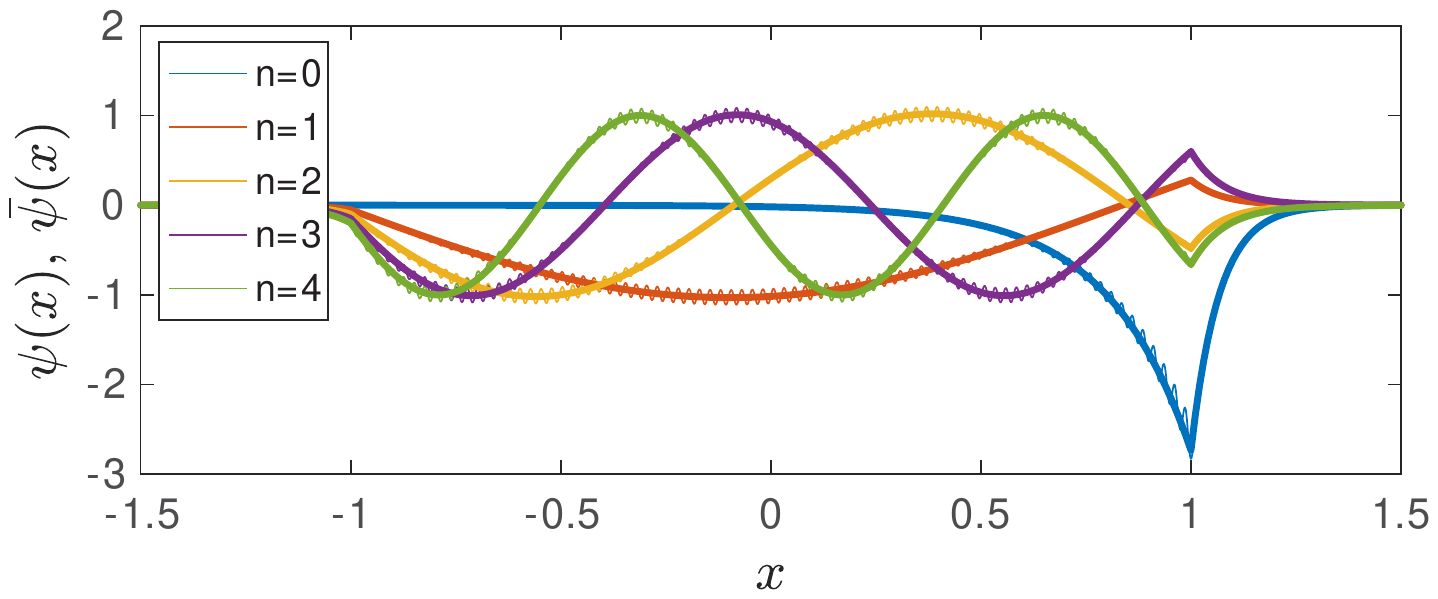}
\caption{\label{fig:pw_delta} (Color online) The case of piecewise continuous
  envelope~(\ref{pw_envel}). The five lowest-energy states calculated
  numerically using the original Hamiltonian~(\ref{main}) (thin oscillating
  curves) and found semi-analytically for the effective
  potential~(\ref{pw_pot1}) (thick curves). The parameters are set to $a = 20$
  and $k \approx 250$ is chosen in such a way that $g(\varphi_{-1,1})=-1$.}
\end{figure}

In order to demonstrate the asymmetric and singular case, we use $v(s)=\sin(s)$ 
whose antiderivative is $g(s) = -\cos(s)$. 
Again, we set $\varphi_{\pm 1} = 0$, 
but now the effective potential~(\ref{pw_pot1}) has the shape of a square quantum well 
with two extra delta-function peaks situated at the edges: A repulsive (resp.\ attractive) 
delta-function peak of strength $a$ is centered at $x = -1$ (resp.\ $x = 1$). 
Fig.~\ref{fig:pw_delta} again shows an excellent agreement between the exact numerical
calculation and the results obtained using the effective potential. As expected from
Eq.~(\ref{pw_pot1}), the probability density is increased close to the right side of 
the potential well, i.e.\ at the position of the attractive delta-function singularity.

\section{Overlapping laser beams}
\label{sec:application}

The preceding analysis shows that the case of oscillating potentials with a zero mean 
is much more intriguing than that with a finite mean. In this subsection, we describe a 
practical setup that can be straightforwardly realized for ultracold atoms moving in 
optical lattices. 

\begin{figure}
\centering\includegraphics[width=85mm]{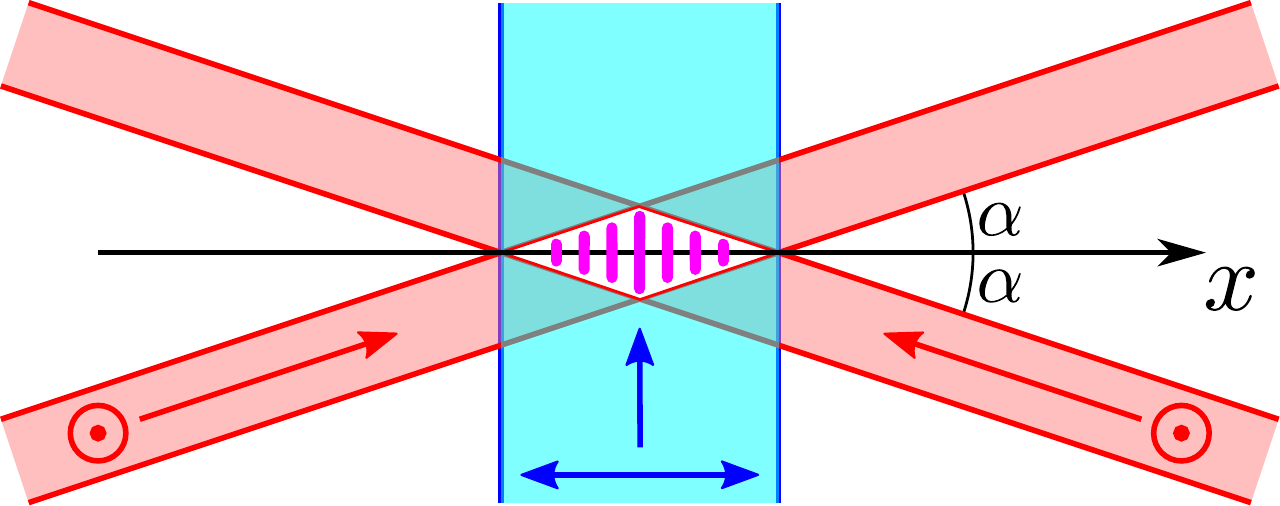}
\caption{\label{fig:lasers} (Color online) Laser-beam configuration for the creation 
of an oscillating potential with zero mean. Two red-detuned beams intersect at an angle
and create an interference pattern. An additional blue-detuned beam is polarized in 
an orthogonal direction and contributes an additional repulsive profile.}
\end{figure}

We start with two coherent laser beams, red-detuned from the atomic resonance, polarized 
in the same (e.g., vertical) direction and intersecting at an acute angle $2\alpha$, as 
shown in Fig.~\ref{fig:lasers}. We assume that the beams are characterized by the angular 
frequency $\omega$, the wavenumber $\kappa$, and the cross-sectional intensity profile 
is described by a Gaussian function of width $b$, i.e.,
\begin{equation}
\label{eq:}
  I = I_0 \exp \left( - \frac{r_{\perp}^2}{2b^2}\right),
\end{equation}
here, $r_{\perp}$ is distance of a given point from the central axis of the beam.
Let us consider the setup where an external trapping potential restricts the motion 
of ultracold atoms in the vicinity of the $x$ axis. Near the $x$ axis
the respective electric fields created by the two beams are given by
%
\begin{subequations}
\begin{align} 
  E_1 &= E_0 \cos (\omega t - \kappa x \cos \alpha) 
    \exp \left[ -\frac{x^2 \sin^2 \alpha}{2b^2} \right], \\
  E_2 &= E_0 \cos (\omega t + \kappa x \cos \alpha) 
    \exp \left[ -\frac{x^2 \sin^2 \alpha}{2b^2} \right].
\end{align}
\end{subequations}
The resulting intensity distribution is given by the time-averaged square of the total 
electric field $E_1 + E_2$. Thus
\begin{equation}
\begin{split}
\label{eq:}
  I &\sim 
  2 E_0^2 \cos^2 (\kappa x \cos \alpha) 
    \exp \left[ -\frac{x^2 \sin^2 \alpha}{b^2} \right] \\
  &= E_0^2 \left\{ 1 + \cos (2 \kappa x \cos \alpha) \right\}
  \exp \left[ -\frac{x^2 \sin^2 \alpha}{b^2} \right],
\end{split}
\end{equation}
and the resulting intensity distribution creates a rapidly oscillating potential 
profile with a slowly-varying Gaussian envelope of a characteristic width $b / \sin\alpha$
which may be much larger than the wavelength. Although this potential does not have
a zero mean, the background contribution 
\begin{equation}
\label{eq:}
  V_{\rm bg} \sim E_0^2 \exp \left[ -\frac{x^2 \sin^2 \alpha}{b^2} \right]
\end{equation}
can be canceled by applying an additional blue-detuned laser beam (see Fig.~\ref{fig:lasers})
polarized in the orthogonal direction and creating a broad Gaussian repulsive potential.

\section{Conclusions and outlook}
\label{sec:concl}

We showed that quantum dynamics in potentials with fast spatial oscillations can be 
approximately described by a smoothed wave function that reproduces the overall
structure of the true wave function but neglects its rapid small-scale oscillations.
The equation of motion for the effective dynamics retains the form of the 
Schr\"{o}dinger equation (\ref{main_eff1}) and can be analyzed based on the usual 
framework and intuition 
available in single-particle quantum mechanics. In particular, generalizations to
few- or many-particle problems can be readily made. In this context, it is intuitively 
clear that inter-particle interactions will not be modified at large length scales,
however, at short scales --- comparable to the range of the performed averaging ---
interesting modifications may take place and are worth to be investigated in future
work.

\begin{acknowledgments}
This research was supported by the Lithuanian Research Council under grant No.~APP-4/2016.
\end{acknowledgments}

\appendix*
\section{Alternative derivation of the effective Hamiltonian}

Here we derive the effective Schr\"{o}dinger equation (\ref{main_eff1}) for quantum 
dynamics in rapidly oscillating spatial potentials using an alternative approach, 
inspired by an analogous treatment of classical motion in the presence of a rapidly 
oscillating force \cite{LandauMech}. 

Our task is to approximately solve the Schr\"{o}dinger equation
\begin{equation}
\label{eq:jr1}
  E \psi (x) = - \frac{\rd^2 }{\rd x^2} \psi (x) 
  + \big[ k v (kx) \Phi (x)  + \gru (x) \big] \psi (x),
\end{equation}
with a slowly varying envelope function $\Phi (x)$ and a rapidly oscillating function
$v (s)$ for a large but finite $k$. Anticipating the result obtained in the main text, we 
represent the wave function as a sum 
\begin{equation}
\label{eq:jr2}
  \psi (x) = \avg{\psi} (x) + \xi (x),
\end{equation}
of the smoothed wave function $\avg{\psi} (x)$ and a correction $\xi(x)$ that scales as $k^{-1}$. 
Substituting the wave function (\ref{eq:jr2}) into the Schr\"{o}dinger equation (\ref{eq:jr1}) 
we obtain 
\begin{equation}
\label{eq:jr3}
  E \avg{\psi} + E \xi = -\frac{\rd^2}{\rd x^2} \avg{\psi} -\frac{\rd^2}{\rd x^2} \xi
  + k \Phi v \avg{\psi} + k \Phi v \xi + \gru \avg{\psi} + \gru \xi,
\end{equation}
which separates into two equations for oscillatory and smooth terms, respectively. 
Focusing first on the oscillating part 
\begin{equation}
\label{eq:jr4}
  E \xi = - \frac{\rd^2}{\rd x^2} \xi + k \Phi v \avg{\psi} 
  + k \Phi v \xi + \gru \xi,
\end{equation}
we collect the terms that are of the order of $k$ 
and arrive at a differential
equation for the correction $\xi (x)$
\begin{equation}
\label{eq:jr5}
  0 = - \frac{\rd^2}{\rd x^2} \xi + k \Phi v \avg{\psi}.
\end{equation}
Assuming that $\Phi$ and $\avg{\psi}$ change slowly, the solution can be written as
\begin{equation}
\label{eq:jr6}
  \xi (x) = k^{-1} \Phi (x) \avg{\psi} (x) w (kx),
\end{equation}
with $w'' (s) = v (s)$; here, $w(s)$ is also a periodic function with zero mean.
Averaging Eq.~(\ref{eq:jr3}) over one spatial period we obtain
\begin{equation}
\label{eq:jr7}
  E \avg{\psi} = - \frac{\rd^2}{\rd x^2} \avg{\psi} 
  + k \Phi \avg{v\xi} + \gru \avg{\psi}.
\end{equation}
Let us evaluate the average $\avg{v\xi}$ in the second term on the right-hand side. 
Using Eq.~(\ref{eq:jr6}) we get
\begin{equation}
\label{eq:jr8}
  \avg{v\xi} = k^{-1} \Phi \avg{\psi} \langle vw \rangle 
  = - k^{-1} \Phi \avg{\psi} \left\langle (w')^2 \right\rangle.
\end{equation}
This leads to the result, equivalent to Eq.~(\ref{main_eff1}) since $w'(s)$ is the 
antiderivative of $v(s)$, which is denoted $g (s)$ in the main text.

\bibliography{fast-spatially-osc}

\end{document}